# CLIMATE STABILITY AND POLICY: A SYNTHESIS

Gerald E. Marsh

Argonne National Laboratory (Ret)
5433 East View Park
Chicago, IL 60615

E-mail: geraldemarsh63@yahoo.com

**Abstract.** During most of the Phanerozoic eon, which began about a half-billion years ago, there were few glacial intervals until the late Pliocene 2.75 million years ago. Beginning at that time, the Earth's climate entered a period of instability with the onset of cyclical ice ages. At first these had a 41,000 year cycle, and about 1 million years ago the period lengthened to 100,000 years, which has continued to the present. Over this period of instability the climate has been extraordinarily sensitive to small forcings, whether due to Milankovitch cycles, solar variations, aerosols, or albedo variations driven by cosmic rays. The current interglacial has lasted for some ten thousand years—about the duration of past interglacials—and serious policy considerations arise as it nears its likely end. It is extremely unlikely that the current rise in carbon dioxide concentration—some 30% since 1750, and projected further increase over the next few decades—will significantly postpone the next glaciation.

**Introduction.**

In trying to understand climate over the Phanerozoic eon (the last 570 million years) one must take into account past high-level concentrations of carbon dioxide linked to the evolution of the Sun as well as paleogeography. Even then large uncertainties remain. Nevertheless, some very simple considerations can lead to a general understanding of climate over this enormous period of time. Before discussing what these are, it is important to get an overall idea of what proxy data tells us of the climate during the Phanerozoic. This can be achieved by a careful examination of the excellent representation from Wikipedia [1] of temperature derived from the available proxy data, shown below as Fig. 1.

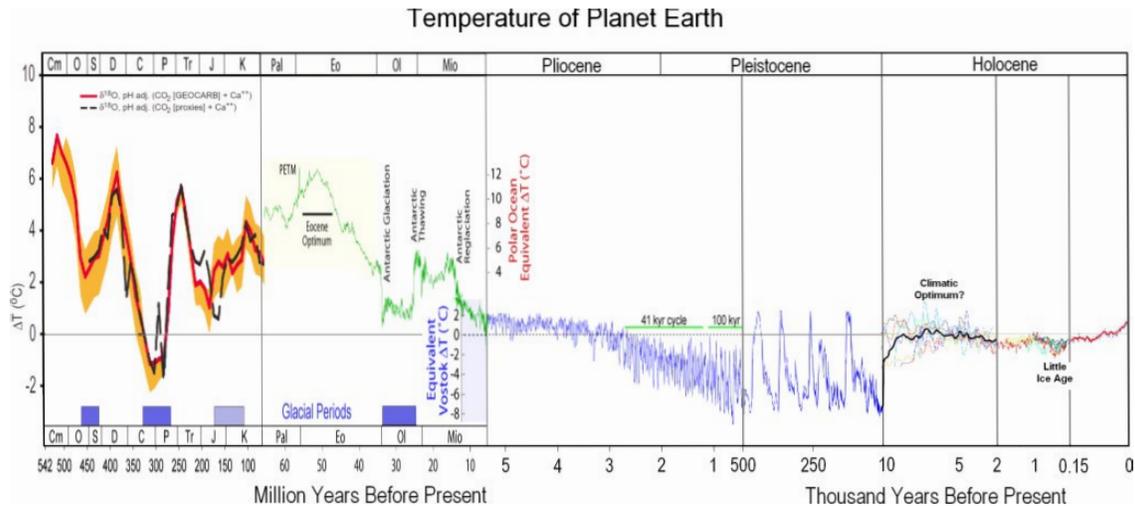

Figure 1. Temperature of the Earth over the Phanerozoic eon. Figure from Wikipedia [1].

As can be seen from the figure, with the exception of the event some 300 Myr ago, the Earth's temperature for most of the Phanerozoic was significantly warmer than it has been over the last five million years. Some 2.75 million years ago, the climate entered an oscillatory phase, initially with a 41 kyr period. For the last million years the spectrum has been dominated by a 100 kyr period. The interglacial intervals are generally considerably shorter than the glacial ones. In other words, the Earth for the last 5 million years has been, on the average, colder than at any time in the last 550 million years, except for a glacial period 300 million years ago. This is despite the increasing luminosity of the Sun over the whole of the Phanerozoic. The reasons for this, as we shall see, have to do with the evolution of life.

The closing of the Isthmus of Panama some 3 million years ago is also thought by many to have contributed to the initiation of the cyclic ice ages. The formation of the isthmus took some 10-15 million years. After its final closure 3 million years ago, it cut off the Atlantic and Pacific oceans forcing a reorganization of the currents between them. This presumably led to the present day thermohaline circulation in the Atlantic, of which the Gulf Stream is a part. Somewhat counter-intuitively, it is argued that the Gulf Stream brought greater warmth to the north Atlantic and therefore warmer and wetter weather to northeastern Europe that contributed to the formation of the Arctic ice cap.

There are serious uncertainties in this scenario. The Gulf Stream is really a boundary current that is part of a large subtropical circular current system called a gyre. It is driven by the tropical trade winds, which blow from east to west, and the mid-latitude westerlies, blowing in the opposite direction. Winter climate in Europe is moderated, north of say 35 degrees latitude, primarily by atmospheric circulation—it does not require a dynamical ocean. The Gulf Stream is responsible for warming Europe by only a few degrees [22]. Therefore, the forming of the Isthmus of Panama is unlikely to be the principal cause of the unstable climate of the past 3 million years.

**Solar evolution and carbon dioxide concentrations.**

The sun is a main-sequence star whose luminosity in the past ($t \leq t_0$) is given by the well-founded expression [2]

$$L(t) = \left[1 + \frac{2}{5}\left(1 - \frac{t}{t_0}\right)\right]^{-1} L_0,$$

where $L_0$ is the current luminosity of the sun and $t_0$ its present age. When the sun first began to shine 4.6 billion years ago, $t$, and consequently $t/t_0$, was zero and today $t/t_0 = 1$. Using the Stefan-Boltzmann relation to determine the equilibrium temperature of the Earth, this expression tells us is that, after the Earth's formation some 4 billion years ago, the temperature of its surface would have been below the freezing point of water for the first 2 billion years of its existence [3]. But this is impossible since there are sedimentary rocks, which can only form in liquid water, dating back some 3.8 billion years. Indeed, the early Earth was warmer than today since there was no glaciation prior to about 2.7 billion years ago.

The most likely explanation for a warm early Earth is the carbon dioxide geochemical cycle. Without liquid water, weathering of silicate rocks would cease, and carbon dioxide released by volcanoes would accumulate in the atmosphere until the temperature was above freezing, allowing weathering to resume. While carbon dioxide is a minor greenhouse gas, large changes in its concentration can significantly alter climate.

Assuming that silicate rocks can be represented by the mineral wollastonite ($CaSiO_3$), the chemistry behind the weathering process is as follows [4]:

$$CaSiO_3 + 2\,CO_2 + H_2O \rightarrow Ca^{++} + 2\,HCO_3^- + SiO_2$$

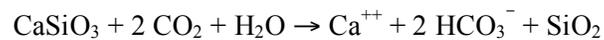

$$Ca^{++} + 2\,HCO_3^- \rightarrow CaCO_3 + CO_2 + H_2O$$

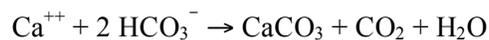

The net result being,

$$CaSiO_3 + CO_2 \rightarrow CaCO_3 + SiO_2.$$

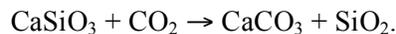

Similar reactions hold for Mg silicate weathering. These reactions summarize many intermediate steps. The inverse reaction, back to silicate rocks, is driven by metamorphism and magmatism (see the references for details).

How much carbon dioxide would be needed to compensate for the dimmer Sun in the past can be obtained by solving the equation

$$242\left(\cfrac{1}{1+\cfrac{2}{5}\left(1-\cfrac{t}{t_0}\right)}-1\right)+\alpha\ln(\mathrm{RCO2})=0$$

for RCO2, the ratio of the carbon dioxide concentration in the past to its present value. Here, 242 w/m² is the net radiation absorbed by the Earth—averaged over the whole Earth—taking into account the amount reflected due to the Earth's albedo,[*] and $t_0$ is 4.6 X 10⁹ years. The result, converting RCO2 to ppmv using the pre-industrial concentration for carbon dioxide of 280 ppmv, is shown in Fig. 2.

It is difficult to know what value would be best to choose for $\alpha$ given the large variability of this coefficient in the publications of the Intergovernmental Panel on Climate Change (IPCC) and in the literature. Berner, et al. [5], use for the product of $\alpha$ and the climate sensitivity $\lambda$ the value 2.88 °C. For the generally accepted value for climate sensitivity of $\lambda = 0.53$ °C w⁻¹m², this corresponds to a value of $\alpha$ of 5.4 w/m². This is the value that will be used here. The value $\alpha\lambda = 2.88$ °C used by Berner corresponds to a temperature rise of 1.9 °C for a doubling of carbon dioxide concentration, a value that is not inconsistent with past IPCC reports and with the work of Royer, *et al.* [6], who find that "a ΔT(2X) of at least 1.5 °C has been a robust feature of the Earth's climate system over the past 420 Myr."

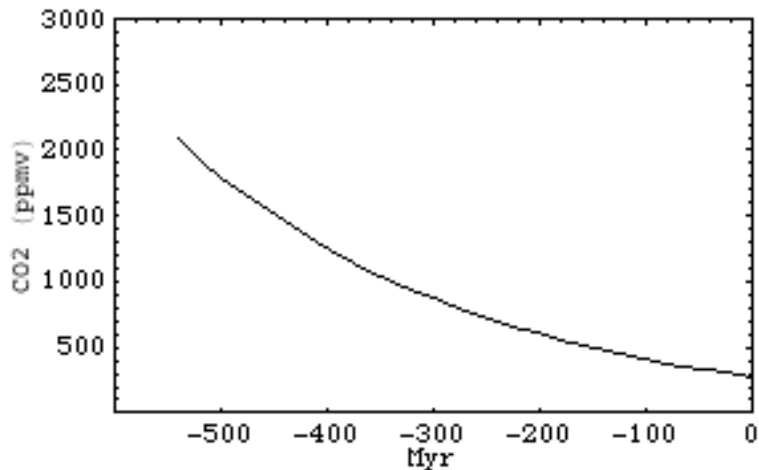

Figure 2. The concentration of carbon dioxide needed to compensate for the dimness of the Sun over the Phanerozoic assuming a pre-industrial (1750) concentration of 280 ppmv. Note the lack of linearity.

At the beginning of the Phanerozoic some 570 Myr ago, the Sun was about 4.5% dimmer than today. If the radiative forcing of this reduced solar output was compensated by greater carbon dioxide levels, the concentration had to be some 7.5 times the 1750 value of 280 ppmv. To give an intuitive feel for what a change in luminosity of 4.5% means, solar-type stars have luminosity variations of 0.1% to 0.4% [7]. Between the twelfth century Medieval Maximum and the Maunder Minimum of 1645-1715, the brightness of the Sun is estimated to have decreased by 0.5% [8]. This was enough to cause the Earth to enter the Little Ice Age, illustrating the extraordinary sensitivity of modern climate to small variations in forcing.

---

[*] Of course, to be in radiative equilibrium, the Earth must also radiate this amount into space.

A reconstruction of Phanerozoic carbon dioxide levels has been given by Berner and Kothavala [9] using the GEOCARB III model. The results are shown in Fig. 3. Note that the most extensive and long lasting glaciations are the current cycle of ice ages and the glaciation that occurred during the Permian-Carboniferous period centered about 300 Myr ago. One cannot look at this figure without asking what caused the carbon dioxide concentration to precipitously drop prior to 300 Myr ago followed by its subsequent rise and fall until the present.

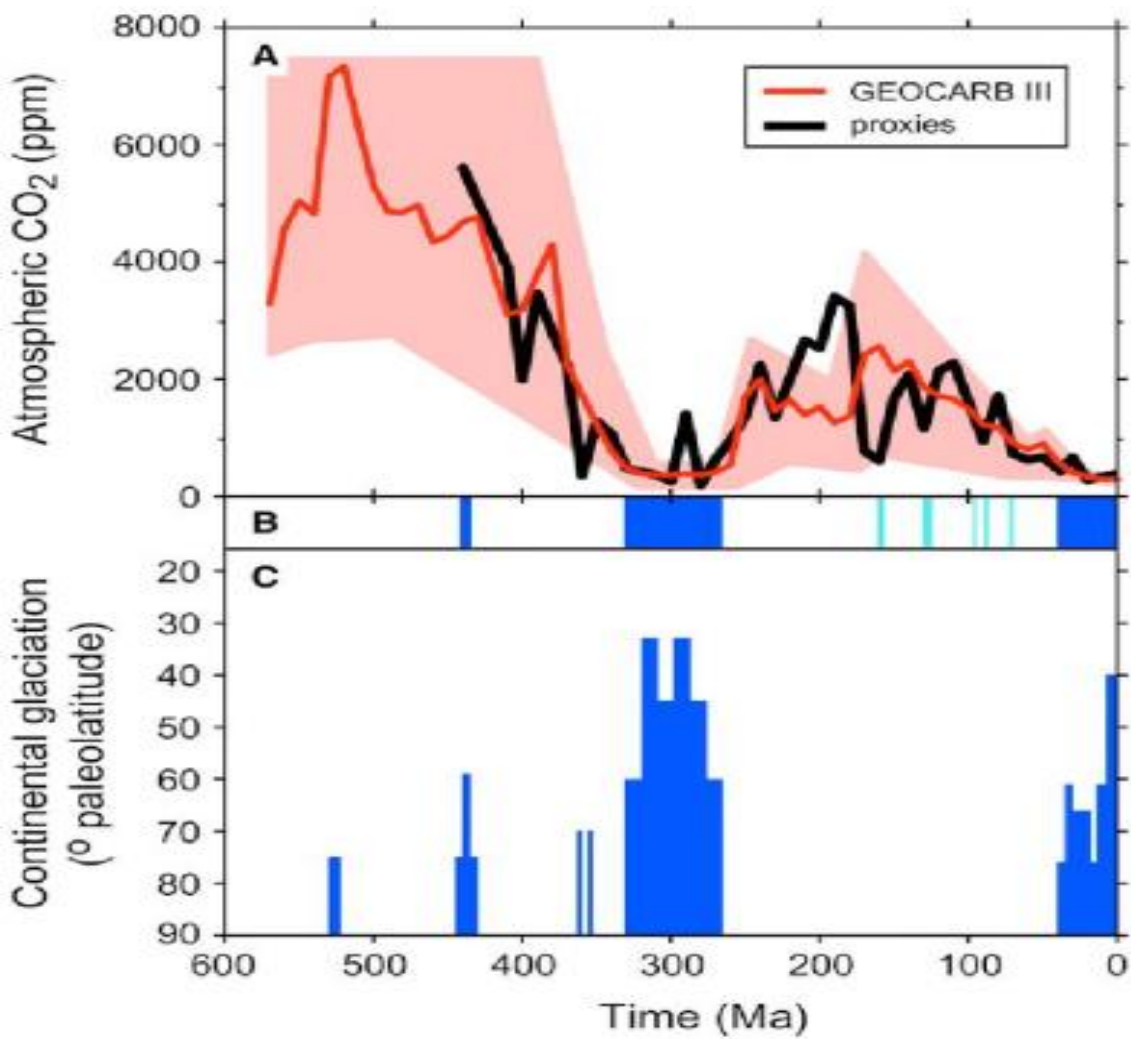

Figure 3. The correlation between carbon dioxide and climate: A, GEOCARB III and proxy reconstruction of carbon dioxide concentrations. The shaded area gives an idea of the error range [9]; B, glacial and cool climates; C, latitude extent of glaciation [10]. (Figure from Berner and Kothavala [9].)

On a multi-million year time scale, the carbon dioxide removed from the atmosphere by weathering must be in balance with the amount released by degassing (the inverse of the weathering reactions given above). Because weathering is primarily a function of temperature, assuming a constant rate of degassing, one would expect a gradual drop in carbon dioxide concentration over the Phanerozoic as the Sun's increasing luminosity caused an increase in temperature and weathering. The relatively sudden drop beginning around 400 Myr ago has to represent some dramatic change that was taking place on the surface of the Earth at the time. And there was one. The drop in carbon dioxide during the Permian-Carboniferous was coincident with the rise of large vascular land plants, which greatly increased the weathering of silicate rocks [11]. This is because land plants enhance silicate-weathering rates, *inter alia*, by increasing the partial pressure of carbon dioxide in soils by a factor of 10 to 40 over the atmospheric value.

Because an increase in atmospheric carbon dioxide concentration raises temperature, which in turn increases weathering, weathering acts as a negative feedback response to increasing carbon dioxide. Thus, increased weathering—due to an increase in the partial pressure of carbon dioxide in soils from land plants—reduces the Earth's surface temperature by removing carbon dioxide from the atmosphere, consequently slowing the weathering process. It is believed that this is the mechanism that gradually restored the balance between temperature and weathering after the rise of large vascular land plants.

The spread of large vascular plants over the land also increased the burial of organic material in sediments. These plants are composed of about 25% lignin, which together with cellulose forms their woody cell walls. At the time of their rise, few microbes capable of decomposing lignin existed, and as a result large amounts of this material were buried during the Carboniferous and Permian, forming the raw material for the Earth's vast coal reserves.

That the climate became cold some 300 Myr ago should not be viewed as contradicting the impression one usually has of the Carboniferous being a time of lush vegetation and large insects—due to elevated oxygen levels—since at the time the continents were in different locations and formed a super continent known as Pangaea, much of it being located in tropical latitudes.

The dip in the concentration of carbon dioxide that reached its minimum some 300 Myr ago should be viewed as a large perturbation in the long-term decrease in carbon dioxide over the whole of the Phanerozoic due to the increasing luminosity of the Sun, which—by raising temperatures—increased the weathering rate and slowly lowered carbon dioxide levels.

An obvious way to check all this is to ask what the temperature and carbon dioxide concentration at the Earth's surface would be if land plants were eliminated today, and what the concentration was before the rise of large vascular land plants. A rough approximation of what the impact on carbon dioxide levels would be if land plants were eliminated has been given by Kasting and Grinspoon [2]. The correction factor $f_w$ to the present-day silicate weathering flux due to a change in carbon dioxide concentration in the atmosphere is given as a function of the resulting temperature difference by

$$f_w = 1 + 0.087(T - T_0) + 0.0019(T - T_0)^2,$$

where $T_0$ is 288 °K (the present mean surface temperature). The weathering "flux" is given by the product of the first order rate constant for weathering, the correction factor $f_w$, and the amount of silicate material undergoing weathering.

Berner, et al. [5] gave this equation in terms of atmospheric carbon dioxide content, but since weathering is primarily a function of temperature Kasting and Grinspoon have written it with temperature as the independent variable. As mentioned above, the weathering and degassing cycle is assumed to be in long-term balance so that $f_w = 1$ for the modern Earth where $T = T_0$. The rise in temperature due to an increase in the partial pressure of carbon dioxide is given by Berner, et al. as

$$T - T_0 = 2.88 \ln(P/P_0),$$

where $P_0$ and $T_0$ are respectively the initial partial pressure and temperature. The number 2.88 is the product $\lambda \alpha$ discussed above.

Thus far, weathering has been assumed to be independent of carbon dioxide concentrations, but while this made sense in the context of the model developed by Berner, et al., this assumption needs to be modified in the present context. Since land plants increase the partial pressure of carbon dioxide ($pCO_2$) in soils by a factor of 10-40, this factor needs to be accounted for when considering the effect of their elimination. Kasting and Grinspoon assume existing data can be extrapolated so that the weathering rate varies approximately as $(pCO_2)^{0.3}$ for conditions of interest. To maximize the climatic effect, they also assume that the elimination of land plants would reduce soil $pCO_2$ by a factor of 40. The correction factor $f_w$ must then take account of the quantity $[P_s/40P_0]^{0.3}$, where $P_s$ is the partial pressure in the soil. On the modern Earth, $P_s = 40P_0$ so this factor is unity. One can therefore approximately include the effect of land plants on the weathering factor $f_w$ by writing it as

$$f_w = [P_s/40P_0]^{0.3} + 0.087(T - T_0) + 0.0019(T - T_0)^2.$$

Note that this formula again yields $f_w = 1$ for $T = T_0$ and $P_s = 40P_0$ thereby showing that the weathering-degassing cycle is in balance for the modern earth. If land plants are eliminated, $P_s \rightarrow P_0$, and the first term on the right hand side of this equation becomes $[1/40]^{0.3} = 0.29$. Imposing the requirement that the carbon cycle be in balance, so that $f_w = 1$, yields the equation

$$(T - T_0)^2 + 45.79(T - T_0) - 373.68 = 0.$$

To obtain $P/P_0$, one substitutes the equation given above for $T - T_0$ into this equation and solves the following quadratic in $\ln(P/P_0)$ for $P/P_0$:

$$[\ln(P/P_0)]^2 + 16 \ln(P/P_0) - 45 = 0.$$

The result is $P/P_0 = 11.5$. Using $P_0 = 280$ ppmv, this corresponds to a carbon dioxide concentration of about 3220 ppmv, corresponding to a temperature rise of about 7 °C.

To see what the carbon dioxide concentration would have been hundreds of millions of years in the past, however—before the rise of large vascular land plants—one must take the reduced solar luminosity into account. These plants began to populate the land some 400 Myr ago, so from Fig. 2 the concentration of carbon dioxide to be added is some 1250 ppmv. Adding this to the 3220 ppmv derived above results in a concentration of 4470 ppmv, or an increase over pre-industrial (1750) values by about a factor of 16, which is in good agreement with GEOCARB III, as can be seen from Fig. 3.

Over the Phanerozoic, carbon dioxide is believed to be the principal driver of climate [11], although variations in cosmic ray flux due to passing through the spiral arms of the galaxy [12] or rising above the galactic plane [13] could play a secondary role on a multi-million year time scale.

Something should be said about the glacial period of around 445 Myr ago, which while short—lasting some two million years—was quite extensive. At this time, carbon dioxide concentrations were quite high (GEOCARB III predicts levels of about 4200 ppmv), and yet there was glaciation, challenging the apparent correlation between carbon dioxide and temperature. Given the lower luminosity of the Sun at the time, the carbon dioxide threshold for initiating glaciation may have been considerably higher than today [14]. Nevertheless, the causes of this glaciation remain somewhat mysterious. Some have advanced the idea that an increase in albedo due to cosmic rays could be responsible, although this remains controversial. But an increased albedo due to other causes is also a possibility.

The above has only touched the surface of the complex carbon dioxide geochemical cycle and carbonate metamorphism. Nevertheless, the broad outline should be clear: the evolution of the Earth's climate over the Phanerozoic was driven by the slow increase in the luminosity of the Sun and the carbon dioxide geochemical cycle. Silicate weathering served as a negative feedback mechanism, countering increased carbon dioxide concentrations so that a balance was struck. This feedback was modified by the biota, and in particular the rise of large vascular land plants, further reducing carbon dioxide levels.

**Net forcing due to carbon dioxide and solar luminosity changes.**

It is interesting, and important because of the policy implications, to examine the net forcing over the Phanerozoic and compare the result with the timing of the glaciations. This can be done by using the expression given above for the luminosity of the sun and adding to it the forcing due to carbon dioxide concentrations as given by GEOCARB III data and the proxy data shown in Fig. 3. The result is shown in Fig. 4.

As can be seen from the figure, net radiative forcing is negative (weaker than at present) only during two periods, the long-lived and extensive glaciation of the Permo-Carboniferous centered about 300 Myr ago, and the last 3 million years or so of the late Cenozoic. The resolution of the data is not adequate to show relatively short-lived inter-glacials, if they existed, for the former period. For more on this, see the discussion in [14].

In Fig. 1, the cyclical ice ages of the late Cenozoic display a 41-kyr cycle that transitioned to a 100-kyr cycle about 1 million years ago. If this behavior, unstable compared to much of the Earth's history over the Phanerozoic, is to be driven by variations in insolation caused by periodic changes in the Earth's orbital precession, eccentricity, and obliquity—that is, by the Milankovitch cycles, the phasing of the ice ages and orbital forcing need to correspond. At first glance, this does not seem to be the case. In particular, the 100 kyr spectral signature does not appear to match the frequencies associated with the Milankovitch cycles.

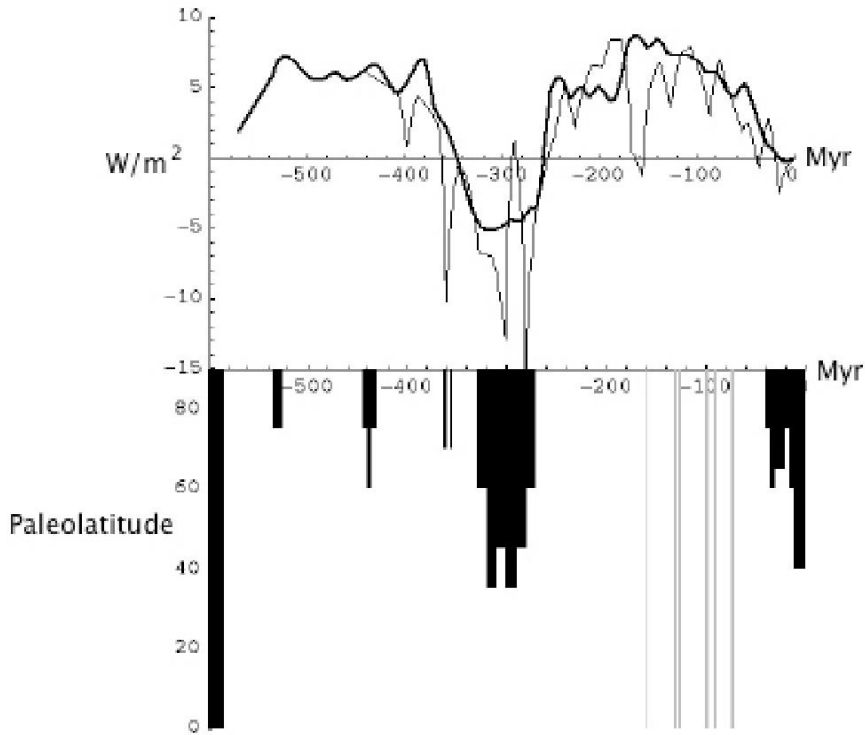

Figure 4. Net forcing due to variation of the luminosity of the Sun and carbon dioxide concentrations as given by GEOCARB III data (heavy curve) and the proxy data (light curve) over the Phanerozoic. The calculation takes account of the Earth's albedo of about 29%. The shaded lines centered around 100 Myr in the past correspond to cool periods unreferenced to paleolatitude.

Various explanations for this have been given in the literature: Ehrlich details the problems with the Milankowitch theory and proposes that resonant thermal diffusion waves in the sun could be responsible for the 100 kyr cycle [15]; Muller and MacDonald have argued that the 100 kyr signature is due to variation in the inclination of the Earth's orbit relative to the invariant plane of the solar system coupled with the presence of interplanetary dust [16]; and Ridgwell, Watson, and Raymo argue against any such mechanism and suggest that Milankovitch cycles give an adequate explanation if deglaciations only occur every fourth or fifth precessional cycle [17].

The details of these possibilities don't matter for the purpose of this paper, only the observation that each of these explanations involves only small forcings, showing again the extraordinary sensitivity of the Earth's climate system over the late Cenozoic.

**Climate sensitivity.**

Roe and Baker [18], in looking at the potential long-term increase in mean global temperature in response to a doubling of carbon dioxide concentration, showed that the probability distributions associated with such projections are relatively insensitive to decreases in the uncertainties associated with the underlying climate processes. The approach they used was the standard feedback analysis employed for many purposes including electronic feedback-amplifier theory. The same methodology will be used here to look at the response of climate to a decrease in solar irradiance comparable to that of the Little Ice Age (LIA).

For a change in radiative forcing, the equilibrium change in global temperature, $\Delta T$, is $\Delta T = \lambda \Delta R_f$, where $\lambda$ is the climate sensitivity and $R_f$ is the change in radiative forcing, which—for the case being considered here—could be due to a change in solar irradiance or the Earth's albedo. In the absence of feedback processes, it is generally assumed that the reference climate sensitivity is $\lambda_0 = 0.3\ °Cw^{-1}m^2$.

The best data available on total solar irradiance from 1600 to 2000 were given in 2001 by the IPCC in Fig. 6.5 of their report *Climate Change 2001: The Scientific Basis*. They gave the change in solar irradiance between the LIA and around 1850 (after the LIA) as about 1.75 w/m$^2$. The 2007 IPCC report rescaled this data by a factor of 0.27 based on the work of Yang, et al. [19]. The figure from the 2001 IPCC report is shown below.

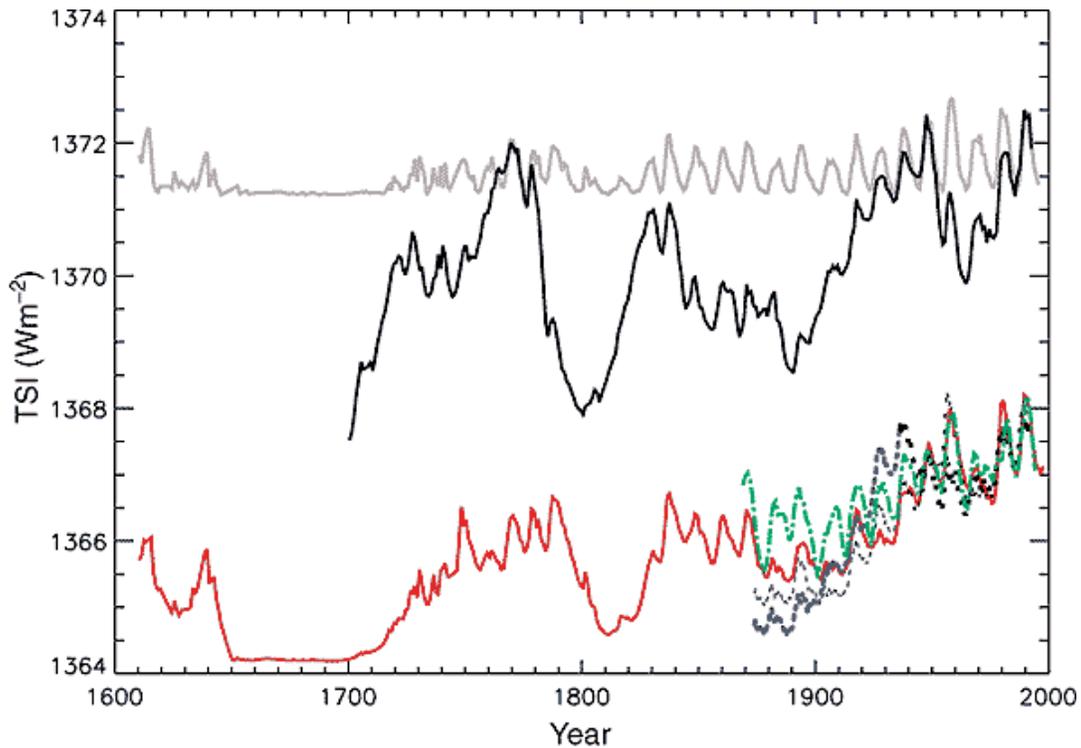

Figure 5: Reconstructions of total solar irradiance (TSI) by Lean et al. (1995, solid red curve), Hoyt and Schatten (1993, data updated by the authors to 1999, solid black curve), Solanki and Fligge (1998, dotted blue curves), and Lockwood and Stamper (1999, heavy dash-dot green curve); the grey curve shows group sunspot numbers (Hoyt and Schatten, 1998) scaled to Nimbus-7 observations for 1979 to 1993. [Fig. 6.5 and caption from *Climate Change 2001: The Scientific Basis*] (Color on-line)

The updated estimate by Yang, et al., gives a *model-dependent* average increase in total solar irradiance from the Maunder minimum (the time of the LIA) to an average around 1850 as being about 0.7 w/m$^2$. Using this value in the methodology developed below, however, yields unreasonable values for the total climate feedback in response to a change of solar irradiance. For this reason, an intermediate value of 1 w/m$^2$ will be used here. This, more conservative approach, reduces the sensitivity of climate to changes in solar forcing. A decrease in solar irradiance of 1 w/m$^2$ corresponds to a decrease in solar forcing of 0.178 w/m$^2$.

The equilibrium change in temperature, $\Delta T_0$, due to a change in solar irradiance of 1 w/m$^2$ is then $\Delta T_0 = \lambda_0 \Delta R_f = (0.3 \, ^\circ\text{Cw}^{-1}\text{m}^2) \times (0.177 \, \text{w/m}^2) = 0.053 \, ^\circ\text{C}$. This is without any feedbacks from the climate system. Such feedbacks will affect the forcing, which in turn modifies $\Delta T$. Along with Roe and Baker, it is assumed here that the functional relation is $\Delta T = \lambda_0(\Delta R_f + c \, \Delta T)$, where $c$ is a constant. Let the total feedback factor, including feedbacks from multiple underlying climate processes, be defined as $f = \lambda_0 c$. Then one may express the latter functional relation as

$$\lambda = \frac{\Delta T}{\Delta R_f} = \frac{\lambda_0}{1-f}.$$

A model-independent estimate of the climate sensitivity, including all feedbacks, to a change in solar irradiance can be calculated from data from the LIA. This in turn allows the feedback factor $f$ to be calculated from the above formula.

The average global reduction in temperature during the LIA is generally accepted to be about 0.4 $^\circ$C. If the reduction in solar irradiance for the LIA is 1 w/m$^2$, the change in forcing as given above is 0.178 w/m$^2$, and therefore the climate sensitivity, including all feedbacks, is

$$\frac{\Delta T}{\Delta R_f} = \frac{0.4 \, ^\circ\text{C}}{0.178 \, \frac{w}{m^2}} = 2.25 \, ^\circ\text{Cw}^{-1}m^2.$$

Using the previous equation, this gives a value for $f$ of $f = 0.867$. If the rescaled change in solar irradiance of 0.7 w/m$^2$ were used, the result would be $f = 0.9$; alternatively, if the original un-rescaled data were used from Fig. 5, corresponding to a change in solar irradiance of 1.75 w/m$^2$ for the period of interest, the resulting feedback would be 0.77. These are large feedback values compared to the mean value for carbon dioxide given by Roe and Baker as 0.65 ($0.42 \leq f \leq 0.73$).

Such a large feedback factor goes a long way towards explaining the extraordinary sensitivity of the climate system to small changes in forcing due to changes in solar irradiance, albedo, or insolation changes caused by Milankovitch cycles.

There are many uncertainties in the various feedbacks that make up the total feedback factor $f$. The effects of these uncertainties, following Roe and Baker, will be assumed to result in a normal distribution for $f$. Its average value will be assumed here to be $\bar{f} = 0.867$ as determined above, and the standard deviation of $f$ will be chosen to be $\sigma_f = 0.13$, typical—according to Roe and Baker—of feedback studies using global climate models.

The change in temperature as a function of $f$, given the equilibrium change in temperature, $\Delta T_0 = 0.053$ °C due to a change in solar irradiance of 1 w/m², is then

$$\Delta T(f) = \frac{\Delta T_0}{1 - f}.$$

As $f \to 1$, the system approaches an unstable regime. For a decrease in forcing, $\Delta T_0$ is negative, and consequently so is $\Delta T$. Because $\Delta T$ is not a linear function of $f$, the distribution for $\Delta T$ which, using Roe and Baker's notation is $h_T(\Delta T)$, is not normal but is obtained in the following way.

The normal distribution for $f$ is given by

$$h_f(f) = \frac{1}{\sigma_f \sqrt{2\pi}} \exp\left[-\frac{1}{2}\left(\frac{f - \bar{f}}{\sigma_f}\right)^2\right].$$

Now $f$ can be viewed as a function of $\Delta T$, that is, $f = f(\Delta T)$. Taking the derivative of the expression above for $\Delta T$, and multiplying the resulting equation by $h_f(f)$ allows one to write

$$h_f(f(\Delta T)) \frac{df(\Delta T)}{d(\Delta T)} = h_f(f(\Delta T)) \frac{\Delta T_0}{(\Delta T)^2} = h_T(\Delta T).$$

$h_T(\Delta T)$ is defined by the quantity on its left. Note that $h_T(\Delta T)$ as defined has the property that as $f \to 0$ or $1$, $h_T(\Delta T) \to 0$. Since, from the above,

$$f(\Delta T) = 1 - \frac{\Delta T_0}{\Delta T},$$

the distribution $h_T(\Delta T)$ can be written

$$h_T(\Delta T) = h_f\left(1 - \frac{\Delta T_0}{\Delta T}\right) \frac{\Delta T_0}{(\Delta T)^2}.$$

Using the expression for the normal distribution $h_f(f)$ given above, $h_T(\Delta T)$ takes its final form

$$h_T(\Delta T) = \frac{1}{\sigma_f \sqrt{2\pi}} \frac{\Delta T_0}{(\Delta T)^2} \exp\left[-\frac{1}{2}\left(\frac{\left(1 - \frac{\Delta T_0}{\Delta T}\right) - \bar{f}}{\sigma_f}\right)^2\right].$$

The distributions and their relationships are shown below in Figure 6.

Note that the probable error for the feedback factor, $P.E.$—defined such that 50% of the data falls between $\bar{f} \pm P.E.$, is given by $P.E. = 0.6745\ \sigma_f = 0.0877$. Added to $\bar{f}$ this gives 0.95, perilously close to unity.

As pointed out by Roe and Baker, "The basic shape of $h_T(\Delta T)$ is not an artifact of the analyses or choice of model parameters. It is an inevitable consequence of a system in which the net feedbacks are substantially positive." The long tail of the skewed distribution $h_T(\Delta T)$ means that there is a not insignificant probability of large changes in temperature in response to relatively small changes in forcing. Keep in mind that the difference between the LIA and current global temperatures is only about 1.1 °C.

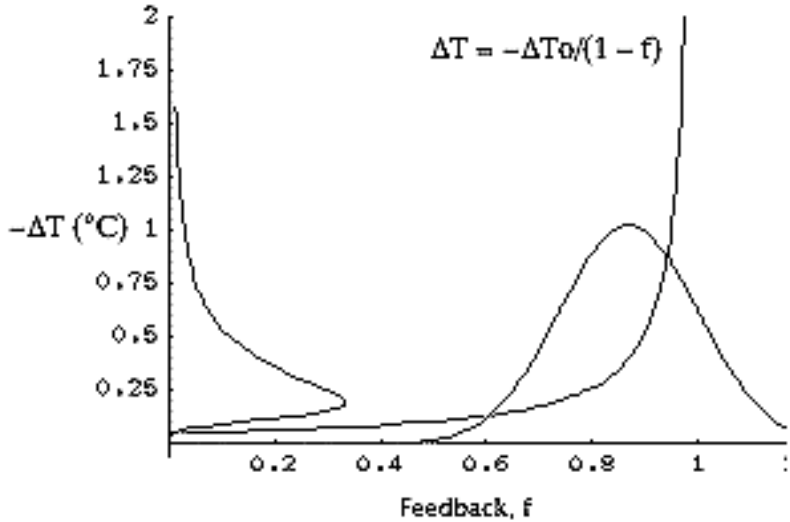

Figure 6. The normal distribution $h_f(f)$ is shown on the right of the figure, while the distribution $h_T(\Delta T)$ is on the left. The ordinate, showing temperature, is to be associated with the distribution $h_T(\Delta T)$, while the abscissa, showing the feedback factor $f$, with $h_f(f)$. The third curve shows $\Delta T$ as a function of $f$. The temperature difference $-\Delta T_0$, excludes all feedbacks and corresponds to a decrease in temperature of 0.053 °C due to a drop in solar forcing comparable to the Little Ice Age. The average feedback $f = 0.867$ intersects the curve for $\Delta T(f)$ at a temperature of 0.398 °C. Note that even a $1\sigma$ positive deviation from the mean in $f$ gives a feedback very close to unity.

**Policy Implications.**

It is shown above that the carbon dioxide geochemical cycle coupled with the evolution of both the Sun and biota over the Phanerozoic has led to the exceptionally low value of atmospheric carbon dioxide concentration that characterizes modern times. These low levels have in turn resulted in the Earth entering a period of instability characterized by the cyclical ice ages of the past 2.75 million years. The present extraordinary sensitivity of climate to small changes in forcing, whether due to Milankovitch cycles affecting insolation, solar variations as occurred during the Little Ice Age, variation in stratospheric aerosols, or cosmic ray driven albedo variations, is a result of the low carbon dioxide concentrations that have remained generally below 500 ppmv beginning some 20 million years ago. Although proxy data show concentrations of this gas occasionally falling below this level previous to 20 million years ago, the average was above [14]. The glacial period centered around 300 Myr in the past was perhaps an exception.

The current inter-glacial period has lasted for some ten thousand years, comparable to the length of past inter-glacials. While policy considerations over the last couple of decades have concentrated on potential effects of rising temperatures—due, it is believed by many, to the increase in carbon dioxide concentrations from anthropogenic sources—these concentrations are quite low relative to those during times of climate stability that include most of the Phanerozoic. Even if all the temperature increase over the last century is attributable to human activities, a doubtful proposition at best, the rise has been a relatively modest 0.7 $^o$C, a value within natural variations over the last few millennia. While an enduring temperature rise of similar magnitude over the next century would cause humanity to face some changes that would undoubtedly be within our spectrum of adaptability, entering a new ice age would be catastrophic for the preservation of modern civilization. One has only to look at maps showing the extent of the glaciation during the last ice age to understand what a return to ice age conditions would mean. Even if the transition took centuries, the historical records of the Little Ice Age make it clear that life would become increasingly difficult even in the early stages.

Over the near term, NASA maintains that Solar Cycle 25, peaking around 2022, could be one of the weakest in the last three centuries [20]. The sunspot minima around this time will be comparable to the Dalton Minima around 1805, and could cause a very significant cooling (see Fig. 5 and compare to the Maunder Minimum of 1645-1715, the time of the Little Ice Age).

There has been much speculation in both the scientific and popular literature that increased warming as a consequence of anthropogenic carbon dioxide emissions could lead to an increased flow of fresh water into the north Atlantic that would shut down the thermohaline circulation, known alternately as the meridional overturning circulation or the Atlantic heat conveyor [21]. This in turn it is argued, could initiate a new ice age in Europe. There are two major misconceptions behind such speculation: First, the Gulf Stream is not responsible for the transport of most of the heat that gives Europe its mild climate [22]; and while the shut down of the thermohaline circulation does appear to play an important role in the dramatic drop in temperature due to Heinrich and Dansgaard-Oeschger events [23], such shutdowns can only occur during an ice age. Indeed, Broecker [24], who first linked the thermohaline circulation to the ice ages, now discounts the fear that a shutdown of the thermohaline circulation could trigger an ice age. He has pointed out that for that scenario to work feedback amplification from extensive sea ice is required [25]. The possibility that global warming could trigger an ice age through shutdown of the thermohaline circulation may therefore be discounted.

Given that the real danger facing humanity is a return to a new ice age, it makes sense to ask what concentration of carbon dioxide would be adequate to stabilize climate so as to extend the current inter-glacial indefinitely. Some idea of the range of concentrations needed can be had from the work of Royer [14] who found that over the Phanerozoic consistent levels of carbon dioxide below 500 ppmv are associated with the two glaciations of greatest duration—those that occurred during the Permo-Carboniferous some 300 Myr ago and the Cenozoic, within which we are now living. Cool climates were found to be associated with carbon dioxide concentrations below 1000 ppmv, while no cool periods were associated with concentrations above 1000 ppmv.

Some support for the idea that moderately increased carbon dioxide concentrations could extend the current interglacial period comes from the work of Berger and Lautre [26]. Working with projections of June insolation at 65 °N as affected by Milankovitch variations over the coming 130 kyr, they used a 2-dimensional climate model to show that moderately increased carbon dioxide concentrations, coupled with the small amplitude of future variations in insolation, could extend the current interglacial by some 50 kyr. The insolation variations expected over the next 50 kyr are exceptionally small and occur only infrequently, the last time being some 400 kyr in the past. They also found that a carbon dioxide concentration of 750 ppmv would *not* extend the interglacial beyond the next 50 kyr. In addition, concentrations of less than 220 ppmv would terminate the current interglacial.

One should not, however, take these carbon dioxide concentrations as the last word. The sensitivity of the climate to a doubling of carbon dioxide concentration could be in error. Since 1990, estimates by the IPCC of the coefficient $\alpha$, discussed earlier, vary by 15-19% and "implicitly include the radiative effects of global mean cloud cover" [27], and estimates of the radiative effect of clouds are quite uncertain. If the actual sensitivity is significantly lower than current estimates, that would elevate the concentration of carbon dioxide needed to extend the current interglacial.

IPCC projections for carbon dioxide concentrations by the year 2100 depend on projections of social and industrial development in countries with large populations that currently consume small amounts of energy per capita. The highest concentrations projected are about 1100 ppmv. This projection could be exceeded, however, if development in China and India accelerates and if other underdeveloped nations are able to overcome current impediments to modernization.

Even if development continues along its current trajectory, carbon dioxide concentrations are almost certain to fall in the range of 500-1000 ppmv over the next century. This is because there are very good reasons to be pessimistic about current approaches to limiting carbon dioxide emissions—they are simply not realistic, instead being the result of political rather than scientific considerations. This is an observation, not a criticism since the current approach may be the best that is possible given existing international relationships and law, along with other aspects of political reality.

Two examples regarding fossil fuels may suffice to illustrate realistic constraints on curtailment of their use. First consider oil. Its use in industry is widespread for a variety of purposes in addition to energy production, but it will be irreplaceable in the transportation sector for decades. Apart from niche applications for other fuels, there are simply no good alternatives that are economically and politically viable. Some may be tempted to believe that the use of oil will be self-limiting, forcing the use of alternative fuels. This point of view is based on the claims of "peak oil" theorists. Such claims, however, show a misunderstanding of the meaning of "oil reserves". These reserves depend on price and are not a direct measure of the amount of oil physically available in the ground. There is plenty of oil, perhaps as much as the 7200 billion barrels estimated by ExxonMobil, but these reserves cannot be brought to market as cheaply as oil from the Persian Gulf, and the economics of oil dictates that cheaper oil will be used first. Moreover, these sources cannot begin production immediately; there is a ramp up period of years. If the phasing-in of such reserves does not match the decline of current oilfields, rising prices and conflict over resources are inevitable. In the end the oil will become available.

The argument that biofuels could replace oil is worth discussing. While the substitution of biofuels in the transportation sector appears at first blush promising, it has the severe handicap of competing with food production. Extensive development without careful planning is likely to raise the cost of food and other agricultural products much more than it already has. Nor is it clear how planning could be done without interfering with the market mechanisms needed for efficient production—existing subsidies have already had this effect.

There are other problems. One attractive choice for biodiesel fuel is rapeseed oil, but to produce enough biodiesel from this source to fuel the country would require some 1.4 *billion* acres. For comparison, the U.S. now has only 400 million acres under cultivation. In addition, there is the fresh water, already in short supply, and the fertilizer needed for this increased cultivation. Even if cellulose can be used as a feedstock, biofuels based on agriculture are unlikely to replace oil any time soon.

Another example is electricity. In the United States, about 40% of the carbon dioxide emissions are from the burning of fossil fuels to generate electricity. Projections by the International Energy Agency and the Energy Information Administration indicate that alternative sources of electricity such as solar and wind have no possibility of being able to displace this use of fossil fuels any time soon, if ever. The choice is between coal and nuclear, and the latter, while currently undergoing a limited renaissance, is beset by political obstacles, one of which is the prevalent concern about waste disposal. This concern, however, is also political [28].

Nevertheless, there is only one practical way known today to stabilizing carbon dioxide concentrations over the next few centuries: nuclear power coupled with the long-term development of a hydrogen economy based on nuclear energy. A hydrogen economy does not necessarily mean that nuclear generated hydrogen is burned directly; the hydrogen may be used in the production of liquid fuels, should it turn out that such fuels are the most efficient and economical means for storage and distribution. But other than the current feeble attempts to implement a Global Nuclear Energy Partnership, this is not even on the international agenda.

Unless the international approach to stabilizing carbon dioxide concentrations changes dramatically, the world will continue to depend on fossil fuels for generations to come, and the burning of such vast quantities of fossil fuels is bound to have a serious environmental impact. The developed world cannot legislate how the developing world will use these fuels, and history has shown that commercialization will likely be at the lowest cost to the producer with the concomitant release of vast quantities of pollutants as well as carbon dioxide. China is a perfect contemporary example. Yet if the grinding poverty that most people in the developing world must live under today is to end through development along the Western model—and no alternative model has been shown to be viable—the required energy has to come from somewhere.

Resolving these issues is far beyond the purview of the IPCC. But that United Nations organization could have an important role in the future. The IPCC and the climatology community in general should devote far more effort to determining the optimal range of carbon dioxide concentrations that will stabilize the climate, and extend the current interglacial period indefinitely.

**Acknowledgement.**
I would like to thank George S. Stanford for numerous suggestions to improve the readability of this paper.